\documentclass[10pt,aps,prd,twocolumn,showpacs,superscriptaddress,longbibliography]{revtex4-1} 
\usepackage{graphicx}  
\usepackage{dcolumn}   
\usepackage{bm}        
\usepackage{amssymb}   
\usepackage{amsmath}
\usepackage{mathrsfs}
\usepackage{amsthm}
\usepackage{amsfonts}
\usepackage{oldgerm}
\usepackage{url}
\usepackage{natbib}
\usepackage[utf8]{inputenc}
\usepackage[usenames]{color}
\usepackage[bookmarks=false]{hyperref}

\providecommand \enquote  [1]{#1}%

\def\ss{\scriptscriptstyle}

\newcommand{\UESC}{\affiliation{Laboratório de Astrofísica Teórica e Observacional,
Departamento de \\ Ciências Exatas e Tecnológicas, Universidade Estadual de Santa
Cruz,\\ Rodovia Jorge Amado, km 16, 45662-900 Ilhéus, Bahia, Brazil}}
\newcommand{\UFABC}{\affiliation{Centro de Ci\^encias Naturais e Humanas, Universidade
Federal
do ABC,  Avenida dos Estados 5001, 09210-580 Santo Andr\'e, São Paulo, Brazil}}

\begin{document}

\title{New results on the physical interpretation of black-brane gravitational perturbations}

\author{Enesson S. de Oliveira}\UFABC\UESC

\author{Alex S. Miranda}\UESC

\author {Vilson T. Zanchin}\UFABC

\date{\today}

\begin{abstract}
The linear perturbation theory applied to the study of black holes is a traditional 
and powerful tool to investigate
some of the basic properties of these objects, such as the stability of the event horizon,
the spectra of quasinormal modes, the scattering and the production of waves in a process of
gravitational collapse. Since long ago, the physical interpretation of the
linear fluctuations in the metric of spherically symmetric black holes has been established. 
In a multipolar expansion, it is known that polar perturbations of a monopole type ($l=0$) can only increase the
black-hole mass, axial perturbations of a dipole type ($l=1$) induce a slow rotation in the
system, and perturbations with $l\geq 2$ always lead to the production of gravitational waves.
However, in relation to the planar Schwarzschild anti--de Sitter black holes (or black branes, for short),
there is still no conclusive study on some aspects of the physical meaning of these
perturbations. In particular, there is some controversy concerning the polar sector
of fluctuations with zero wave numbers $(k=0)$. Some authors claim that this kind of
perturbation causes only a small variation in the black-brane mass parameter, while others
obtained also evidence for the existence of gravitational waves associated to such modes.
The present study aims to contribute to the resolution of this controversy by 
revealing the physical meaning of the gravitational perturbations of anti--de Sitter 
black branes. In this work we use the Chandrasekhar's gauge formalism to evaluate the
linear variations in the complex Weyl scalars in terms of the Regge-Wheeler-Zerilli gauge-invariant
quantities. Then we use the Szekeres' proposal for the meaning of the Weyl scalars and the 
Pirani's criterion for the existence of gravitational radiation in order to give a
physical interpretation of the black-brane perturbations with arbitrary wave number values.

\end{abstract}


\maketitle

\section{Introduction}

The perturbation theory is an important tool in studying some of the properties of black holes,
for instance, in investigating the stability of event horizons under metric fluctuations, 
as well as in the study of generation, absorption and scattering of gravitational waves by such objects
(see, e.g., Refs.~\cite{Tagoshi-atall97, Sasaki-Tagoshi2003,Poisson-atall95}).
Furthermore, from the perspective of the AdS/CFT correspondence
\cite{Maldacena99,Witten1998,Gubser-atall1998,Aharony-atall2000},
black holes in asymptotically anti--de Sitter (AdS) spacetimes are dual to a thermal equilibrium state
of a boundary conformal field theory (CFT), and the first-order gravitational perturbations of these
black holes correspond to linear fluctuations in the energy-momentum tensor of the
dual thermal state \cite{Horowitz-Hubeny2000}.

There are two main methods to study gravitational perturbations of black holes.
The first one consists in considering perturbations of the black-hole metric and then by
linearizing the Einstein equations around the given background spacetime. The second method
consists in the linearization of the equations resulting from
the use of the Newman-Penrose
formalism \cite{Newman-Penrose}. In both cases, a major difficulty is writing the 
physically interesting variables in terms of gauge-invariant quantities.
The gauge freedom arises from the identification of events in the background with events in the physical (perturbed) spacetime.
For this reason, the gravitational perturbations are usually described in terms of gauge-dependent
variables, whose gauge freedom may be explored to simplify the analysis, and only the final 
results are written in terms of gauge-invariant quantities.

As far as we know, the first work to present a complete and clear formulation of
the gravitational perturbation theory was published in 1974 by Stewart and Walker \cite{Stewart-Walker74}. 
In the meantime, important progress on the study of metric perturbations of black-hole
spacetimes was made in the work by Regge and Wheeler \cite{Regge-Wheeler57} published in 1957,
where the stability of a Schwarzschild black hole under axial perturbations was studied.  
The gauge used in this paper is now known as the Regge-Wheeler gauge.
Based on the same idea, Zerilli \cite{Zerilli70a,Zerilli70b} analyzed
the gravitational radiation that arises when stars fall into a static black hole and extended
the Regge-Wheeler study for polar perturbations, completing the fundamental equations for the gravitational
perturbations of Schwarzschild black holes. Later on, Chandrasekhar \cite{Chandrasekhar75} obtained
the same equations as Regge-Wheeler and Zerilli by
using a different gauge, that is currently known as the Chandrasekhar gauge. 
Following different routes, Teukolsky \cite{Teukolsky72,Teukolsky73} and Moncrief \cite{Moncrief74a, Moncrief74b} 
developed studies on the gravitational perturbations of the Kerr and Reissner-Nordström black holes, respectively,
with the aim of investigate the stability
of these black-hole spacetimes. 

In more recent years, Kodama, Ishibashi and Seto \cite{KISS2000} developed an interesting strategy
to investigate the gravitational perturbations in static higher-dimensional spacetimes with some
specified spatial symmetry. In such a strategy, the spatial symmetry of the
background spacetime is used from the beginning to decompose the perturbations in three independent sectors, namely,
tensor, vector and scalar sectors, and to construct gauge-invariant quantities for each one of the sectors.
This construction has been used to study several properties of gravitational perturbations in black-hole
and black-brane spacetimes, as for instance in the work by Dias and Reall \cite{Dias20132}, where the algebraically
special modes of higher dimensional black holes are investigated. 
The vector and scalar sectors of the Kodama-Ishibashi-Seto gauge-invariant formulation correspond, respectively, to the 
axial and polar perturbations in the language of the Regge-Wheeler-Zerilli formalism, while the tensor
perturbations are not present in $4$-dimensional spacetimes.

An interesting approach regarding the physical interpretation of a given background metric
and its gravitational perturbations is based on the effects of the curvature
tensor on the relative motion of free test particles through the geodesic deviation equation.
Using a tetrad of null vectors, Szekeres \cite{Szekeres} wrote the geodesic deviation 
equations for empty spacetimes in terms of the Weyl scalars and obtained a physical 
interpretation for these scalars. Motivated by this technique, Podolsk\'y and \v{S}varc
\cite{Podolsky-Svarc2012} arrived at a similar interpretation for the Weyl scalars in
higher-dimensional spacetimes. They went beyond the free part of the gravitational field,
and took into account the isotropic action of the cosmological constant and the influence of matter
in the spacetime curvature. From the Szekeres' and Podolsk\'y-\v{S}varc's
interpretation, it is possible to extract the effects of gravitational perturbations on
freely falling test particles. In particular, it is possible to verify whether a given class of metric
perturbations are really associated to gravitational waves or not. 
The aforementioned approaches allowed to establish a classification and an interpretation of gravitational
perturbations of spherically symmetric black holes, even in the presence of a cosmological
constant. On the other hand, in relation to the case of AdS black-brane perturbations, there still exist
some open questions regarding the physical interpretation of the polar-sector perturbations
with zero wave numbers. For instance, Kodama and Ishibashi \cite{Kodama-Ishibashi2003} argue that
these perturbations produce only a small change in the mass parameter of the black brane, in complete
agreement with the case of a monopole-type ($l=0$) perturbation of a Schwarzschild black hole.
However, in a more detailed study with the use of the Chandrasekhar diagonal gauge,
it was shown in Ref.~\cite{Miranda-Zanchin2007} that the zero wave number polar perturbations  
may also represent gravitational waves along the radial direction, i.e.,
they describe also gravitational waves propagating in the perpendicular direction to the black-brane horizon.

Therefore, motivated by the absence of a conclusive study on this theme, the present paper aims to investigate
the physical meaning of the gravitational perturbations of black branes in asymptotically anti--de Sitter spacetimes
and, consequently, to analyze the possibility of obtaining solutions with gravitational waves for polar perturbations
with zero wave numbers. For this, the perturbations in the Weyl scalars are calculated and, from the Szekeres \cite{Szekeres} approach for the analysis of the geodesic deviation equations, the physical meaning of the perturbations shall
be obtained in an invariant way.
As an additional investigation, the canonical form of the Riemann tensor is calculated for the perturbed black brane, and the Pirani's criterion~\cite{Pirani57} is used to study the physical meaning of the polar  perturbation sector.
In the studies of the metric perturbations related to this problem, the Chandrasekhar gauge is employed. 
Finally, let us stress that the main interest here is in the physical interpretation of the perturbation functions and, in particular, in the zero wave number perturbations. We will not focus on
the calculation of the complete black-brane quasinormal mode spectrum because it is a well studied subject in the literature.
See, e.g., Refs. \cite{Cardoso-Lemos2001,Miranda-Zanchin2006,Berti-Cardoso2009-Review,Morgan-etall2009} for more details.

The structure of this paper is as follows. We present in Sec.~\ref{sec:background} the black-brane
spacetime and define the basic quantities of the Newman-Penrose formalism for such a background.
Section~\ref{sec:pertrubEqs} is devoted to review the gravitational perturbation theory in the
Chandrasekhar gauge, and to present the fundamental equations for the axial and polar metric variations.
In Sec.~\ref{sec:physicalmeaning} we use the Szekeres' proposal to extract the physical meaning of
the gravitational fluctuations with both vanishing and nonvanishing wave numbers. We use in Sec.~\ref{sec:IV}
the Pirani's criterion~\cite{Pirani57} as an alternative way to interpret the polar-sector perturbations,  
in particular the perturbations with zero wave numbers.
In Sec.~\ref{Sec:IVB} we show how the gravitational waves
associated to the zero wave number polar fluctuations arise in the Kodama-Ischibashi-Seto approach.
We conclude in Sec.~\ref{sec:V} by discussing the main results of this paper.

Geometric units are used throughout this text, so that the speed of light $c$ and the gravitational constant $G$ are set to unity, $c= 1=G$.
The signal convention for the Riemann, Ricci, and Einstein tensors is that of Ref.~\cite{Gravitation}. For the Newman-Penrose quantities, the signal convention of Ref.~\cite{Miranda-atall2015} is adopted.

\section{The background spacetime}
\label{sec:background}

The Einstein equations with a negative cosmological
constant admit an asymptotically AdS solution,
whose associated metric can be written in the form \cite{Lemos:1994xp,huang1995,Lemos:1995cm}

\begin{equation}
 ds^2= -f(r,M)dt^2+f^{-1}(r,M)dr^2+r^2(d\varphi^2+dz^2),
 \label{eq:background}
\end{equation}
with
\begin{equation}
  f(r,M)= \frac{r^2}{\ell^2}-\frac{2M}{r},   \label{eq:hfun}
 \end{equation}
where $M$ represents the mass parameter, $\ell^2=-3/ \Lambda_{c}$ is the AdS 
radius, and $\Lambda_{c}$ is the negative cosmological constant.

The local geometry of such a solution is Euclidean in the sense that the surfaces of constant
$t$ and $r$ are locally flat, but the topology can be planar ($\varphi,\, z\in \mathbb{R}$),
cylindrical ($\varphi\in \mathbb{S}^1,\, z\in \mathbb{R}  $), or toroidal ($\varphi,\,z
\in \mathbb{S}^1$). In short, in this text we stick to the geometry and refer
to this solution as an AdS plane-symmetric black hole, or simply as a black brane. 

The zeros of the function $f(r,M)$, given by Eq.~\eqref{eq:hfun}, determine the horizons of the background
spacetime \eqref{eq:background}. The only real root of the equation $f(r,M)=0$ gives the 
location of the event horizon of the black brane,
\begin{equation}
 r_h = (2M\ell^2)^{1/3}.
\label{Horizonte}
 \end{equation}

The present work is partly performed by using the Newman-Penrose formalism \cite{Newman-Penrose}. For this,
we consider a null tetrad basis  $(l^{\mu},\, n^{\mu},\, m^{\mu},\, m^{*\mu})$, where the real null vectors
$l^{\mu}$ and $n^{\mu}$ are, respectively, tangent to the ingoing and outgoing radial null geodesics of
the background solution~\eqref{eq:background}, i.e.,
 \begin{equation}
 l^{\mu} \partial_{\mu}=\frac{1}{f}(\partial_t+f \partial_r), \qquad n^{\mu}  \partial_{\mu} = \frac{1}{2}(\partial_t-f \partial_r ),
 \label{real_vec}
 \end{equation}%
 while the complex null vector $m^{\mu}$ is defined by 
 \begin{equation}
    m^{\mu} \partial_{\mu} =\frac{1}{\sqrt{2} r}(\partial_z+i\partial_{\varphi}),
    \label{complex_vec}
\end{equation}
with the vector $m^{*\mu}$ being the complex conjugate of $m^{\mu}$.

For the above null tetrad, the only nonvanishing
Weyl scalar of the background metric \eqref{eq:background} is
\begin{equation} \label{psi2}
 \Psi_2 =C_{\rho \sigma \mu \nu} l^{\rho} m^{\sigma} m^{*\mu} n^{\nu}= -\frac{M}{r^3}.
\end{equation}
Hence, the black-brane spacetime is type D in the Petrov 
classification.

\section{Gravitational perturbations of black branes}
\label{sec:pertrubEqs}

\subsection{General remarks}

Let us review here the basic properties of the axial (odd) and polar (even)
sectors of the black-brane gravitational perturbations,
since both are important for the present analysis.
The perturbations of the black-brane metric \eqref{eq:background} and its quasinormal modes 
have been extensively investigated in the literature (see, e.g., Refs.
\cite{Cardoso-Lemos2001,Miranda-Zanchin2006,Berti-Cardoso2009-Review,Morgan-etall2009}).   
However, the interpretation of the resulting perturbation fields has
generated some controversy. In order to investigate this problem in more detail,
we start revisiting the gravitational perturbation theory in the Chandrasekhar
gauge formalism \cite{Chandrasekhar75,Chandrasekhar-Friedman1972}.

We denote the components of the background metric \eqref{eq:background} by $g_{\mu \nu}^{\ss{(0)}}$.
In a first-order theory, the gravitational perturbations are defined as the linear variations
$\delta g_{\mu\nu}\equiv h_{\mu\nu}$ on the background metric $g_{\mu\nu}^{\ss{(0)}}$, i.e.,
the perturbed metric is given by $g_{\mu\nu} =  g_{\mu \nu}^{\ss{(0)}} + h_{\mu\nu}$.
In the present case, the nonzero components of the perturbation $h_{\mu \nu}$
may be written as
\begin{equation}
 \begin{aligned}
 & h_{tt} = - 2 f \, \mu_0, &
 \qquad h_{t\varphi} = r^2  q_0, \\
 & h_{rr} =  2 f^{-1}\,\mu_2,   &  \qquad    
 h_{r\varphi} = r^2 q_2, \\
 & h_{zz} =   2r^2 \, \mu_3,    & \qquad   
   h_{z\varphi} = r^2 q_3, \\
 & h_{\varphi \varphi} = 2r^2\psi, &
 \end{aligned}
\label{pertur}
\end{equation}
where the other components of $h_{\mu \nu}$ are set to zero by an 
appropriate gauge choice as done by Chandrasekhar 
\cite{Chandrasekhar75}. The functions $\mu_0$, $\mu_2$, $\mu_3$  
$q_0$, $q_2$, and $q_3$ are all small quantities when compared to unity.

 Under the substitution $\varphi\rightarrow-\varphi$, 
 the variables $q_0$, $q_2$, and $q_3$ induce odd-parity 
variations in the metric, and so they are called axial (or odd) perturbations. 
On the other hand, the variables $\mu_0$, $\mu_2$, $\mu_3$, and  $\psi$ 
induce even-parity variations under sign change of $\varphi$,
and so they represent polar (even) metric perturbations.

In the Chandrasekhar gauge, the Einstein field equations for the gravitational perturbations
result in a set of coupled differential equations.
However, in the case of axially symmetric perturbations, i.e., in the case the perturbation
functions do not depend on the variable 
$\varphi$, the system of equations can be decoupled into two independent sets of equations,
one for each perturbation sector. In fact, due to the plane-symmetric nature of the spacetime, 
it is always possible to choose the orientation of the frame (in the $\varphi,\, z$ plane)
in such a way that the perturbation functions result independent of $\varphi$. 
Hence, without loss of generality, from now on we consider only the axially symmetric perturbations, 
since this choice allows us to study independently each one of the perturbation sectors.  

Since metric \eqref{eq:background} does not depend on the coordinates $t$, $z$, and $\varphi$, 
any perturbation function $F(t,r,z,\varphi)$ in~\eqref{pertur} can be conveniently represented in
terms of Fourier modes as
\begin{equation}
F(t,r,z,\varphi) \sim \widetilde{F}(r) e^{i(m\varphi+kz-\omega t)},    
\end{equation}
with $\widetilde{F}(r)$ being a function of $r$ only, and $m$ and $k$ being real numbers. 
The wave numbers $m$ and $k$ may be quantized on not, depending on the topology of the $t,r$= constant subspace. 
Notice that, since we are interested in axis-symmetric perturbations, we put $m$ to zero.

\subsection{Fundamental perturbation equations}
\label{sec:perturbEqs}

\subsubsection{Axial perturbations}

For nonvanishing wave numbers, it is possible to describe the axial perturbations by a single 
gauge-invariant quantity $Z^{\ss{(-)}}$, defined in 
terms of the quantities 
$\tilde q_2$ and $\tilde q_3$ 
(see, e.g., Ref. \cite{Cardoso-Lemos2001}).
In the Fourier space, this master variable satisfies
the differential equation
\begin{equation}
 \Lambda^2 {\widetilde{Z}}^{\ss{(-)}} = V^{\ss{(-)}}{\widetilde{Z}}^{\ss{(-)}},
 \label{Eqz-}
\end{equation}
where the function ${\widetilde{Z}}^{\ss{(-)}}={\widetilde{Z}}^{\ss{(-)}}(r)$ is defined by
\begin{equation}
{\widetilde{Z}}^{\ss{(-)}} =-rf\left[\frac{d}{d r}\widetilde{q}_{3}(r)-ik {\widetilde{q}}_2(r)\right],   
\label{Defz-}
\end{equation}
the effective potential $V^{\ss{(-)}}$ is given by
\begin{equation}
 V^{\ss{(-)}} = \frac{f}{r^2}\left(k^2-\dfrac{6M}{r}\right),
\label{potV-}
\end{equation}
and $\Lambda^2$ represents the differential operator
\begin{equation}
 \Lambda^2 = \frac{d^2}{dr_*^2} + \omega^2,
\end{equation}
with $r_*$ being the Regge-Wheeler tortoise coordinate, defined in such a way that
$dr_* = f^{-1}dr$. 

For future reference, we introduce here the operators
\begin{equation} \label{Lambdas}
 \Lambda_{\pm} =  \frac{d}{dr_*} \pm i \omega\,,
\end{equation}
which satisfy the relations
$\Lambda^2=\Lambda_{+}\Lambda_{-} =\Lambda_{-}\Lambda_{+}$.

\subsubsection{Polar perturbations}
\label{polar_case}

Similarly to the axial sector, the polar metric perturbations may be combined to construct
a gauge-invariant function $Z^{\ss{(+)}}$, which satisfy the Fourier-transformed
differential equation
\begin{equation}
\Lambda^2 \widetilde{Z}^{\ss{(+)}} = V^{(+)}\widetilde{Z}^{\ss{(+)}},
 \label{EondZ2}
\end{equation}
where, for nonvanishing wave numbers, $\widetilde{Z}^{\ss{(+)}}=\widetilde{Z}^{\ss{(+)}}(r)$ is defined in terms of the
quantities $\widetilde{\psi}$ and $\widetilde{\mu}_{3}$ by
\begin{equation}
 \widetilde{Z}^{\ss{(+)}} =\frac{3Mr}{rk^2+6M}
\left[\left(1+ \frac{rk^2}{3M}\right)\widetilde{\psi}(r)-\widetilde{\mu}_{3}(r) \right],
 \label{Z+}
\end{equation}
and the effective potential $V^{\ss{(+)}}$ is given by 
\begin{eqnarray}
 \hspace* {-.5cm} & V^{\ss{(+)}}= \dfrac{f}{r^2}\left[ k^2 +
 \dfrac{  72M^2\left(M\ell^2 +r^3\right)- 6k^4M\ell^2r} 
{r\,\ell^2(rk^2+6M)^2} \right]\!.&
 \label{Pot+}
\end{eqnarray}
In what follows, we consider the particular case of perturbations
with zero wave numbers.

\subsection{Perturbations with zero wave numbers}
\label{Pert-null}

\subsubsection{General remarks}

The perturbations characterized by zero wave numbers $(m,\,k)=(0,\,0)$ do not
propagate along the directions parallel to the brane. However, they could be associated
to waves propagating along the radial direction. In fact, in this case 
there are additional gauge degrees of freedom and the physical interpretation of
the metric perturbations is not straightforward. Using the Chandrasekhar gauge formalism,
the authors of Ref.~\cite{Miranda-Zanchin2007} have presented a set of solutions for
black-brane perturbations with zero wave numbers. For completeness, and for future reference, we rewrite those solutions here.

\subsubsection{Axial perturbations}
\label{ZeroAxial}

There exists extra gauge freedom that arises in the zero wave number case, i.e., there is no relation among $q_0$, $q_2$ and $q_3$, and so $q_0$ cannot be eliminated. Therefore, it is possible to reduce the metric perturbations of the axial sector
to a single nonzero component, namely, 
\begin{equation}
 h_{t \varphi} = q_0 r^2= -\frac{J}{r},
\end{equation}
where $J$ is a constant. As a consequence, the line element of the
perturbed spacetime reads
\begin{equation}
ds^2 = -f(r,M)dt^2 +\dfrac{dr^2}{f(r,M)} -\frac{2J}{r}d\varphi dt 
+r^2(d\varphi^2 + dz^2).    
\label{MRot}
\end{equation}
Depending on the compactness of $\varphi$ and $z$, the foregoing metric may represent
a slowly ($J$ is small) rotating black string or
black torus \cite{huang1995,Lemos:1994xp}. In the case of a planar topology ($\mathbb{ R}^2$),
a further coordinate transformation (an infinitesimal Lorentz boost) in the $t,\varphi$
plane puts the metric back to the  unperturbed form \eqref{eq:background}, and so the
black brane does not rotate. For the cylindrical and toroidal topologies, 
this Lorentz boost is a globally ``forbidden'' transformation, and then it results 
in a slowly rotating geometry, as just mentioned.

\subsubsection{Polar perturbations}
\label{zerownpolar}

There is a larger variety of phenomena associated to
the zero wave number polar perturbations than the ones associated to the axial modes.
In particular, the gauge freedom of the zero wave number case can be used to write the perturbed metric in the form \cite{Miranda-Zanchin2007}
\begin{equation}
\begin{split}
 ds^2 =& -f(r,M+\delta M)dt^2 +
 f^{-1}(r,M+\delta M)dr^2 \\
 &+ r^2(e^{2\psi}d \varphi^2 + e^{-2 \psi}dz^2),  
 \label{Metri-null}
 \end{split}
\end{equation}
where $\delta M$ stands for an increment in the mass parameter, and 
$\psi(t,r)=\widetilde{\psi}(r)e^{-i\omega t}$ is a perturbation function,
whose Fourier transform satisfies a wave equation of the same
form as Eq.~\eqref{EondZ2}, with $\widetilde{Z}^{\ss{(+)}}=r \widetilde{\psi}$
and the effective potential
\begin{equation}
 V^{(+)} = \frac{2 f}{\ell^2r^2}\left(r^2+\dfrac{M\ell^2}{r} \right).
 \label{Pot+2}
\end{equation}

For the sake of comparison, it is important at this point to investigate the limit $k\rightarrow 0$
of the results presented in Sec. \ref{polar_case}.
In such a limit, Eq.~\eqref{Z+} results in
$\widetilde{Z}^{\ss{(+)}}= r(\widetilde{\psi}-\widetilde{\mu}_{3})/2$.
On the other hand, by comparing metric \eqref{Metri-null} 
with the corresponding relations in \eqref{pertur}, we obtain
$\mu_3=-\psi$ at first order in a perturbative expansion.
Hence, Eq.~\eqref{Z+} reduces to 
\begin{equation}
 \widetilde{Z}^{\ss{(+)}}(r)  = r\, \widetilde{\psi}(r). 
 \label{Z+2}
\end{equation}
This is the same relation between the wave function $Z^{\ss{(+)}}$ and
the metric perturbation $\psi(t,r)$, as it was found in Ref.~\cite{Miranda-Zanchin2007}
for the metric~\eqref{Metri-null}. It is also straightforward verifying that, for a vanishing value of $k$, the potential $V^{(+)}$ given by Eq.~\eqref{Pot+} reduces to the expression given in Eq.~\eqref{Pot+2}. Therefore, we
conclude that the set of equations~\eqref{EondZ2},~\eqref{Z+} and~\eqref{Pot+} describes the polar
perturbations for all wave number values, and hence the zero wave number polar perturbations represent also gravitational waves.  
 
Here it is interesting to compare the zero wave number polar perturbations of planar geometries to the special modes of polar perturbations of spherically symmetric geometries. For spherically symmetric black holes, the physical interpretation of the special polar perturbations with azimuthal number $l=0$ and $l=1$ cannot be obtained from the equations of motion for perturbations with $l\geq 2$. 
In comparison, positive result here introduces a new way to investigate the polar perturbations of black branes with zero wave numbers, just by taking $k=0$ directly into the general equations~\eqref{EondZ2},~\eqref{Z+} and~\eqref{Pot+}.   

To conclude this section let us mention once again that there is a dispute regarding the physical interpretation of the resulting perturbations in the plane symmetric case, since 
some authors conclude that there exist no waves in the zero wave number case of the gravitational perturbations of plane-symmetric AdS black holes
(see  Refs.~\cite{KISS2000,Kodama-Ishibashi2003}). 
In order to investigate this problem more closely, we shall analyze other 
geometric quantities from which the physical interpretation of the
perturbations can be obtained. The Weyl curvature scalars appearing in the
Newman-Penrose formalism are the best candidates
for such an analysis, and so we calculate them below.

\section{The physical meaning of the black-brane perturbations according to the Szekeres' proposal}

\label{sec:physicalmeaning}

\subsection{The physical interpretation of the spacetime curvature}
\label{sec:IIIA}

Here we follow the strategy suggested by Szekeres \cite{Szekeres},
according to which the physical meaning of the Weyl scalars is derived 
from the geodesic deviation equations.  In this proposal,  the
geodesic deviation equations is projected onto an orthonormal basis $(u^{\mu},s^{\mu }, e^{\ \mu}_{\ss{(2)}},
e^{\ \mu}_{\ss{(3)}})$, where $u^{\mu}$ is the four-velocity of the observer and $s^{\mu}$,
$e^{\ \mu}_{\ss{(2)}}$ and $e^{\ \mu}_{\ss{(3)}}$ are orthogonal spacelike four-vectors. The null
tetrad defined in Eqs.~\eqref{real_vec} and \eqref{complex_vec} are related to this new
tetrad by
\begin{equation}
 \begin{aligned}[2]
  l ^{\mu} &= \left(u^{\mu} + s^{\mu}\right),\quad 
  & m^{\mu}= \frac{1}{\sqrt{2}}\left(e^{\ \mu}_{ \ss{(2)}}+i e^{\ \mu} _{\ss{(3)}}\right),\\
 n^{\mu}&=  \frac{1}{2} \left(u^{\mu} - s^{\mu}\right),    \quad
  & m^{*\mu}= \frac{1}{\sqrt{2}}\left(e^{\ \mu}_{\ss{(2)}}-i e^{\ \mu} _{\ss{(3)}}\right).
 \end{aligned}
\label{VecNull}
\end{equation}

On the basis of these relations, Szekeres wrote the geodesic deviation equation in terms of
the Weyl scalars and showed that the scalar $\Psi_4 \; (\Psi_0)$ describes a
gravitational wave propagating in the direction
of $s^\mu$ ($-s^\mu$).
In turn,
the scalar $\Psi_3$ ($\Psi_1$) corresponds to a longitudinal component of the gravitational
field in the direction 
of $s^\mu$ ($-s^\mu$).
Finally, the real part of the scalar
$\Psi_2$ is associated with Newton-Coulombian effects of the gravitational field
with a principal direction $s^\mu$. 

\begin{figure}
 \centering
   \includegraphics[width=1.00\linewidth]{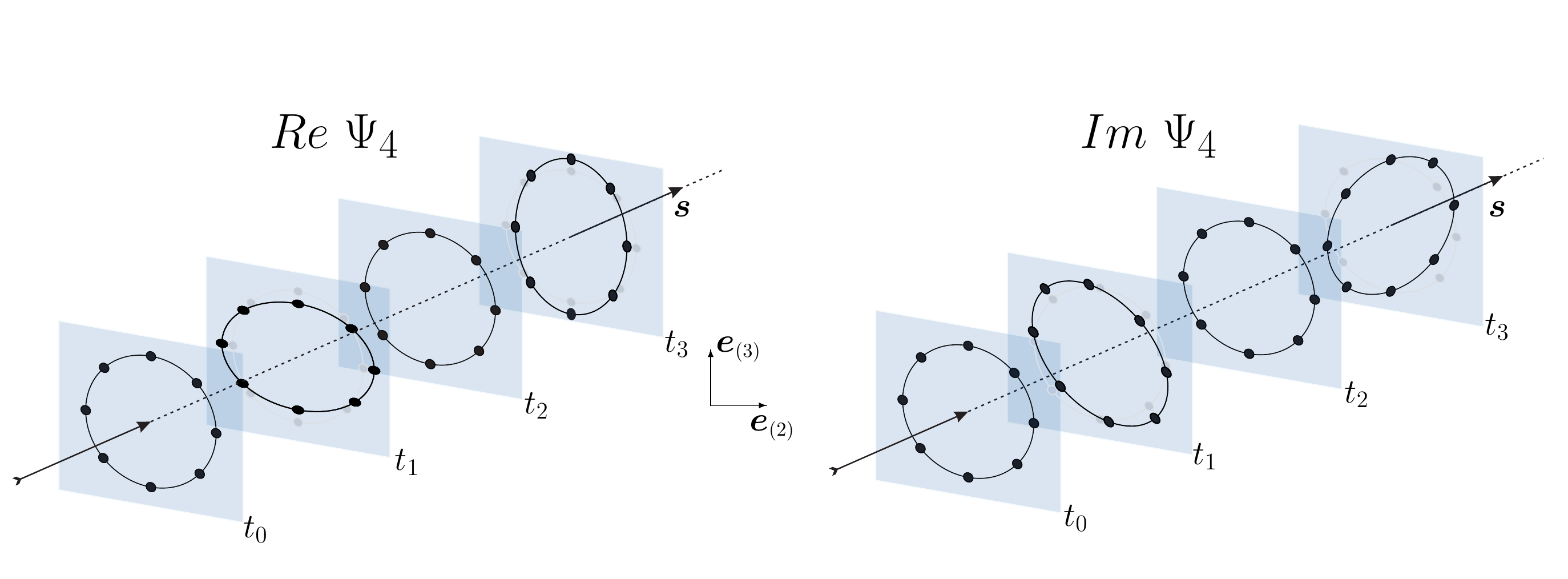}
\caption{The two polarization modes of gravitational waves and the Weyl scalar $\Psi_4$.}
\label{Polarization} 
 \end{figure}

\begin{figure}
 \centering
   \includegraphics[width=0.75\linewidth]{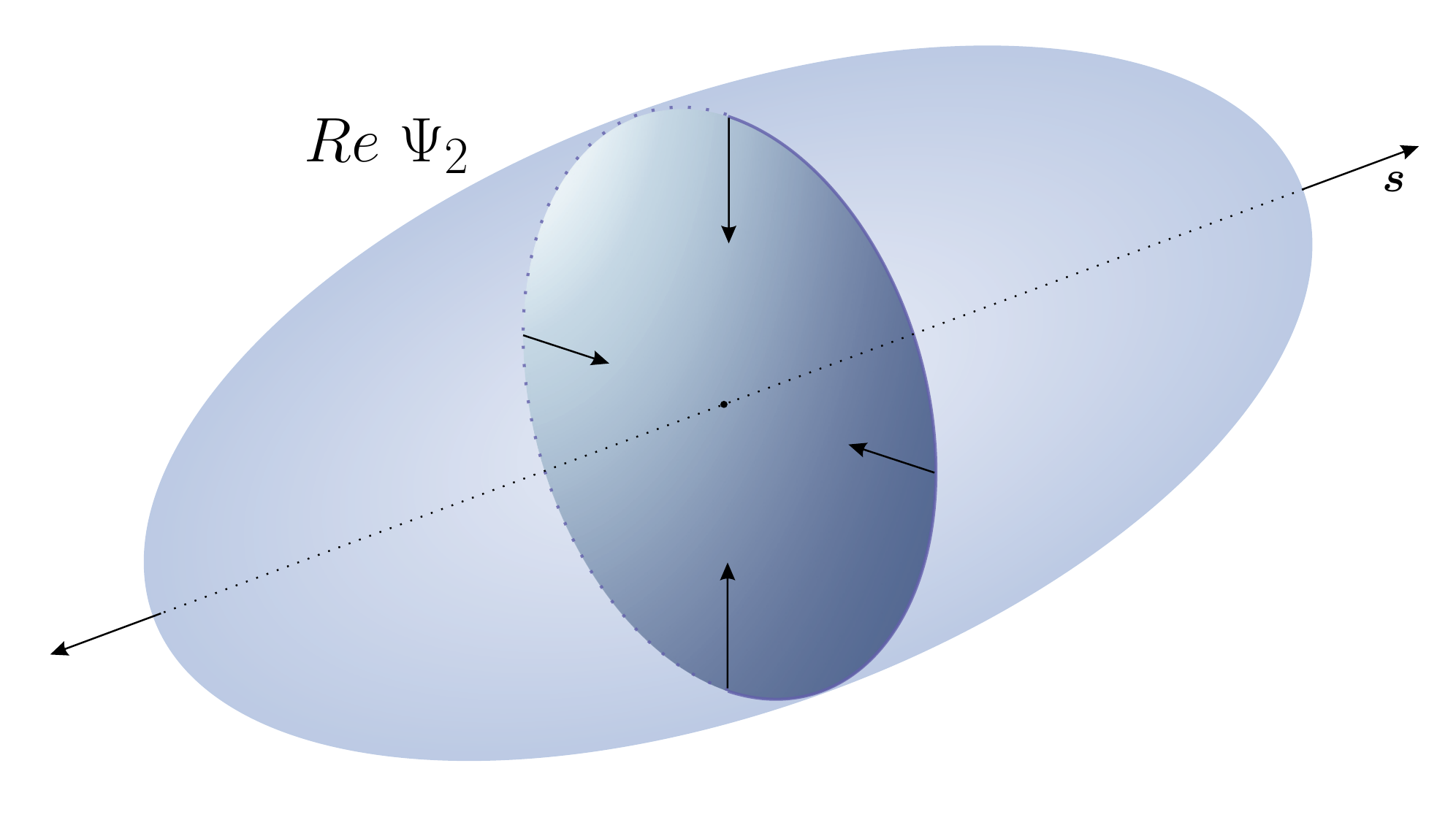}
\caption{Newton-Coulombian effect of $\Psi_2$.}
\label{NC_figure} 
 \end{figure}

It is worth emphasizing that the real and imaginary parts of the Weyl scalar $\Psi_4$ are associated, respectively, with
the ``$+$" and ``$\times$" polarization modes of the gravitational waves propagating in the
$s^\mu$ direction (see Fig. \ref{Polarization}). The two parts of the Weyl scalar $\Psi_0$ produce the same effect that $\Psi_4$ but with gravitational waves propagating in the 
$-s^\mu$ direction. Besides that, just the real part of $\Psi_2$ appears in the geodesic deviation equations.
This part of $\Psi_2$ is associated with a force that deforms a sphere of free particles around an observer, turning it
into an ellipsoid with principal axis in the $s^\mu$ direction 
(see Fig.~\ref{NC_figure}), which is typical of bodies in a central field.

\subsection{Perturbations in a Petrov type-D spacetime}
\label{grav_gauge}

For a general Petrov type-D spacetime, a perturbation $\delta \Psi_{j}$ in an arbitrary Weyl scalar can be written as
\begin{equation}
    \delta \Psi_j= \delta \Psi_j^{\ss{P}}+i\, \delta \Psi_j^{\ss{A}}, \qquad j=0,...,4, 
    \label{separacao_weyl}
\end{equation}
where $\delta \Psi^{\ss{P}}_j$ is the part of the Weyl scalar given in terms of the polar metric perturbations,
while $\delta \Psi^{\ss{A}}_j$ is the part of the Weyl scalar given in terms of the axial metric perturbations.

Moreover, it is important to emphasize that perturbations in the Weyl scalars are
subject to two different kinds of gauge freedom. The first gauge freedom is
connected with the infinitesimal transformations on the tetrad vectors. When a scalar is
invariant under these tetrad transformations, it is said to be tetrad-gauge invariant.
The second gauge freedom is associated to infinitesimal coordinate transformations,
$x^{\mu} \rightarrow x^{\mu}+\xi^{\mu}$, and a quantity that is invariant under this
kind of transformation is said to be coordinate-gauge invariant.

In the case of gravitational perturbations on a Petrov type-D spacetime,
it is possible to show that the perturbations $\delta \Psi_0$ and $\delta \Psi_4$
are both coordinate and tetrad-gauge invariant quantities. However, the perturbations
$\delta \Psi_1$ and $\delta \Psi_3$ are only coordinate-gauge invariant quantities
and, therefore, we can choose a tetrad orientation such that $\delta \Psi_1$ and $\delta \Psi_3$
vanish (see, e.g., \cite{chandrasekhar98book} for more details). Finally, for a Petrov type-D
background, $\delta \Psi_2$ is tetrad-gauge invariant, but just the imaginary
part $\delta \Psi_2^{\, \ss{A}}$ is also coordinate-gauge invariant. For this reason, we are allowed to choose a particular gauge such that $\delta \Psi_2^{\, \ss{P}}=0$.

\subsection{Axial sector}
\label{sec:IIIB.1}

\subsubsection{Perturbations with nonzero wave numbers}

The aim of this subsection is to write the linear variations in the Weyl scalars in
terms of the wave function $\widetilde{Z}^{\ss{(-)}}$ and, from the physical 
interpretation of the Weyl scalars, to extract the meaning of the axial 
gravitational fluctuations of the black branes. In order to obtain an expression
for the Weyl-scalar perturbations in terms of the metric variations, it is necessary to
know the relations between the perturbations in the Weyl scalars and the components of
the Weyl tensor (see, e.g., \cite{Nolan} for details on such relationships).
For the axial perturbations, we write the linear variations of the Weyl scalar
$\delta {\widetilde{\Psi}}_j^{\ss{A}}$
in terms of the metric perturbations~\eqref{pertur} and, after that, we use Eq.~\eqref{Defz-} 
to express the result in terms of the master variable  ${\widetilde{Z}}^{\ss{(-)}}$. After some algebra,
we can then cast the perturbation $\delta {\widetilde{\Psi}}_0^{\ss{A}}$ into the form
\begin{equation}
 2 i\omega\, \delta {\widetilde{\Psi}}_0^{\ss{A}} = \frac{f^{-2}}{r}\left[V^{\ss{(-)}}+
 \left(W^{\ss{(-)}}-2i\omega\right)\Lambda_{-}\right]{\widetilde{Z}}^{\ss{(-)}},
 \label{Psi0Pot3.1}
\end{equation}
where the potential $V^{\ss{(-)}}$ and the operator $\Lambda_-$
are defined respectively by Eqs.~\eqref{potV-} and \eqref{Lambdas},
and the function $W^{\ss{(-)}}$ is given by
\begin{equation}
  W^{\ss{(-)}} = -\frac{6M}{r^2}.
  \label{W-}
\end{equation}
Similarly, we can express the perturbation $\delta {\widetilde{\Psi}}_4^{\ss{A}}$ as
\begin{equation}
2i\omega\, \delta {\widetilde{\Psi}}_4^{\ss{A}} = -\dfrac{1}{4r}\left[V^{\ss{(-)}}+\left(W^{\ss{(-)}} 
+2i\omega\right)\Lambda_{+}\right]{\widetilde{Z}}^{\ss{(-)}},
 \label{Psi4Pot3.1}
\end{equation}
where the notation here is the same as in Eq.~\eqref{Psi0Pot3.1} and the operator
$\Lambda_{+}$ is defined by Eq.~\eqref{Lambdas}.

A necessary and important further step is to write Eqs.~\eqref{Psi0Pot3.1} and~\eqref{Psi4Pot3.1} in terms of the fundamental variables $Y_{\pm 2}$, which arise in the
study of the black-brane gravitational perturbations via the Newman-Penrose formalism
(see, e.g., Ref.~\cite{Miranda-Zanchin2006}). We have that $Y_{+2}= rf^{2}\delta \Psi_0$ and
$Y_{-2}=4r\delta \Psi_4$, and hence, combining these relations to Eqs. \eqref{Psi0Pot3.1} and
\eqref{Psi4Pot3.1}, we obtain
\begin{equation}
 \begin{split}
2i\omega {\widetilde{Y}}_{+2}^{\, \ss{A}} & = 
\left[V^{\ss{(-)}}+(W^{\ss{(-)}}-2i\omega)\Lambda_{-}\right]{\widetilde{Z}}^{\ss{(-)}}, \\
-2i\omega {\widetilde{Y}}_{-2}^{\, \ss{A}} & =\left[V^{\ss{(-)}}+(W^{\ss{(-)}}+2i\omega)\Lambda_{+}\right]
{\widetilde{Z}}^{\ss{(-)}}.
\end{split}
\label{Weyl-nuloY.1}
\end{equation}
From the Szekeres interpretation, we know that the Weyl scalars shown
in Eqs.~\eqref{Psi0Pot3.1} and~\eqref{Psi4Pot3.1} are associated with gravitational waves
propagating in opposite directions. Therefore, we conclude that axial perturbations with $k>0$ generate ingoing and outgoing gravitational waves.  

Lastly, it is possible to write the perturbation $\delta \widetilde{\Psi}_{2}^{\ss{A}}(r)$
in terms of the variable $\widetilde{Z}^{(-)}$ as
\begin{equation}
    \delta\widetilde{\Psi}_{2}^{\ss{A}} = -\frac{k^2}{4r^3}\widetilde{Z}^{(-)}.
    \label{Psi2A}
\end{equation}

\subsubsection{Perturbations with zero wave numbers}

In the case of zero wave numbers, Eq. \eqref{Weyl-nuloY.1} does not hold. Because of extra gauge freedom, the master variable $Z^{\ss{(-)}}$ can be set to zero. Moreover the  scalars $\Psi_0^{\ss{A}}$ and $\Psi_4^{\ss{A}}$ are zero and no gravitational wave is detected. 
As discussed in Sec.~\ref{ZeroAxial}, the axial perturbations with $k=0$ generate at most a slow rotation on the topologically compact black brane spacetimes. The result is the perturbed metric~\eqref{MRot}. 
In fact, for this metric the Weyl scalars  $\delta \Psi_1$ and $\delta \Psi_3$ are proportional to the angular momentum $J$, but as discussed in Sec.~\ref{grav_gauge} the tetrad-gauge freedom may be used to set them to zero. Therefore, axial perturbations with zero wave numbers preserve the type D structure of the background.

\subsection{Polar sector}

\subsubsection{Perturbations with nonzero wave numbers}
\label{sec:IIIB}

In the case of polar perturbations with $k\neq 0$, the Weyl
scalar $\delta \widetilde{\Psi}_{0}^{\ss{P}}$ can be cast into the following form
\begin{equation}
 \delta \widetilde{\Psi}_{0}^{\ss{P}} = -\frac{f^{-2}}{r}\left[V^{\ss{(+)}}+
 \left(W^{\ss{(+)}}-2i\omega\right)\Lambda_-\right]\widetilde{Z}^{\ss{(+)}},
 \label{Psi0Pot3}
\end{equation}
while the scalar $\delta \widetilde{\Psi}_{4}^{\ss{P}}$ can be written as
\begin{equation}
\delta\widetilde{\Psi}_{4}^{\ss{P}} = -\dfrac{1}{4r}\left[V^{\ss{(+)}}+\left(W^{\ss{(+)}} 
+2i\omega\right)\Lambda_+\right]\widetilde{Z}^{\ss{(+)}}.
 \label{Psi4Pot3}
\end{equation}
The function $W^{\ss{(+)}}$ that appears in  Eqs.~\eqref{Psi0Pot3} and~\eqref{Psi4Pot3} is given by
\begin{equation}
  W^{\ss{(+)}} = -\frac{6M(2r^3+k^2\ell^2r+2M\ell^2)}{\ell^2r^2(k^2r+6M)},
  \label{W+}
\end{equation}
and the operators $\Lambda_{\pm}$ are defined in Eq.~\eqref{Lambdas}.

The polar perturbation variables $\widetilde{Y}_{\pm 2}$ can be written in terms of the master variable $ \widetilde{Z}^{\ss{(+)}}$ as 
\begin{equation}
 \begin{split}
\widetilde{Y}_{+2}^{\, \ss{P}} & = 
-[V^{\ss{(+)}}+(W^{\ss{(+)}}-2i\omega)\Lambda_-]\widetilde{Z}^{\ss{(+)}}, \\
\widetilde{Y}_{-2}^{\, \ss{P}} & =-[V^{\ss{(+)}}+(W^{\ss{(+)}}+2i\omega)\Lambda_+]\widetilde{Z}^{\ss{(+)}}.
\end{split}
\label{Weyl-nuloY}
\end{equation}
Equations~\eqref{Weyl-nuloY.1} and~\eqref{Weyl-nuloY} are identical to the results found in
Ref.~\cite{Miranda-Zanchin2006}. It worth mentioning that the authors of Ref.~\cite{Miranda-Zanchin2006} followed a different approach, by using the Newman-Penrose
formalism \cite{Newman-Penrose} and the Chandrasekhar transformation theory~\cite{chandrasekhar98book}, so to get the fundamental Eqs. \eqref{Eqz-} and \eqref{EondZ2}.
Here, however, relations~\eqref{Weyl-nuloY.1} and~\eqref{Weyl-nuloY} were
obtained straightforwardly from the perturbations of the Weyl scalars. 
This fact indicates that both approaches are consistent in the case of  a planar geometry,
just as it happens with gravitational perturbations of spherically symmetric black-hole spacetimes. 

At last, from the Szekeres approach we conclude that zero wave number polar perturbations of black branes 
represent gravitational waves that propagate in two different null directions.

\subsubsection{Perturbations with zero wave numbers}
\label{zeroKpolar}

In the case of polar perturbations with $k=0$, the linear variations 
 $\delta \widetilde{\Psi}_{0}^{\ss{P}}$
and $\delta \widetilde{\Psi}_{4}^{\ss{P}}$ may be reduced
to the same general expressions as given by Eqs.~\eqref{Psi0Pot3} and~\eqref{Psi4Pot3},
now with $V^{\ss{(+)}}$ and $\widetilde{Z}^{\ss{(+)}}$ 
 given, respectively, by Eqs.~\eqref{Pot+2} and~\eqref{Z+2},
while the $W^{\ss{(+)}}$ function now reads
\begin{equation}
W^{\ss{(+)}} = -\frac{2}{r}\left(\frac{r^2}{\ell^2}+\dfrac{M}{r} \right).
\end{equation}
Therefore, as  $\delta \widetilde{\Psi}_{0}^{\ss{P}}$
and $\delta \widetilde{\Psi}_{4}^{\ss{P}}$ are nonvanishing, from the Szekeres interpretation  
we conclude that polar perturbations with zero wave numbers are also associated with 
radial gravitational waves.

It is worth noticing that the wave character of the complete functions $Y_{\pm 2}(t,r)$
may be more easily exhibited through their behavior close to the event horizon 
$r\rightarrow r_h = (2M\ell^2)^{1/3} \ (r_* \rightarrow -\infty)$.
In such a region, the potential $V^{\ss{(+)}}$ vanishes [cf. Eq.~\eqref{Pot+2}]
and then the wave equation \eqref{EondZ2} leads to a solution (in the physical space) of
the form ${Z}^{\ss{(+)}}(t,r) \rightarrow e^{i\omega (\pm  r_*-t)}$. Moreover,
considering the condition of having just ingoing waves at the horizon,
the only allowed solution is ${Z}^{(+)} \rightarrow e^{- i\omega (r_*+t)}$.
As a consequence, we get
\begin{equation}
\begin{gathered}
  \quad \Lambda_{+}{Z}^{(+)} \rightarrow 0,  \quad
\Lambda_-{Z}^{(+)} \rightarrow -2i \omega e^{-i\omega( r_*+t)}, \\
W^{(+)} \rightarrow -\frac{6M}{(2M\ell^2)^{2/3}}\,. 
\end{gathered}
 \end{equation}
Hence, by using the relations~\eqref{Weyl-nuloY}, it is possible to
express the asymptotic form of the functions $Y_{\pm 2}(t,r)$ close to the horizon (for $k=0$) as
\begin{equation}
 \begin{split}
Y_{+2} & \rightarrow 4 \omega \left[\omega -\frac{3Mi}{(2M\ell^2)^{2/3}}
\right] e^{-i\omega( r_* + t)}, \\
Y_{-2} & \rightarrow 0\,.
\end{split}
\label{Y2}
\end{equation}
It is then seen that the function $Y_{+2}$ reduces to a plane wave
traveling radially inwards the black-brane horizon. Additionally, the
function $Y_{-2}$, which is related to the $\delta \Psi_4$ scalar, tends to zero in
such a region. This behavior is expected since, as it was first discussed
in Ref.~\cite{Szekeres}, the scalar $\Psi_4$ represents an outgoing wave
and, because of the imposed boundary condition, it is not possible to
observe such a wave close to the horizon. Moreover, for stationary waves
($\omega = 0$), it also follows that $Y_{+2} =0$, showing no gravitational
radiation at all (as expected). Furthermore, following the same
procedure and considering the outgoing solution ${Z}^{\ss{(+)}} \rightarrow e^{
i\omega(r_*-t)}$, it is possible to show that this solution corresponds to a
plane wave traveling radially outwards.

In short, the conclusion of the analysis presented in this section is that 
the zero wavenumber polar perturbations correspond to
gravitational waves traveling in the radial direction. This, in turn,
answers affirmatively the original question on  whether the
zero wave number gravitational perturbations of black branes can be
associated to the production of gravitational waves, or not.

At last, the computation of the scalar $\delta \Psi_2$ results in
\begin{equation}
 \delta \Psi_2 =-\frac{\delta M}{r^3},
 \label{Weyl-nulo2}
\end{equation}
which describes a perturbation in the ``Coulomb" gravitational field
of the black brane. Such a result is expected from the
discussion presented in Sec.~\ref{zerownpolar}. From  Eq.~\eqref{Metri-null}, 
$\delta M$ is a small variation in the mass parameter of the black brane 
and, therefore, it corresponds only 
to a perturbation in the  Coulomb-type term of the Weyl scalar $\Psi_2$.

\section{The physical interpretation of the polar perturbations from Pirani's criterion}
\label{sec:IV}

\subsection{General remarks}
 \label{sec:IVA}

In accordance with Pirani~\cite{Pirani57}:
{\it ``At any event in empty spacetime, gravitational
radiation is present if the Riemann tensor is of
Type} II {\it or Type} III{\it, but not if it is of Type} I{\it.''}

In the current nomenclature of the Petrov classification, a Riemann tensor of type I represents a spacetime of type I
(non-degenerated Riemann tensor of type I) or a spacetime of type D (degenerated Riemann tensor of type I),
while a Riemann tensor of type II represents a spacetime of type II (non-degenerated Riemann tensor of type II) or a spacetime of type N (degenerated Riemann tensor of type II). Finally, a Riemann tensor of type III corresponds to a spacetime of type III in the Petrov classification.

Another important point to be mentioned here concerns the application of the Pirani's
criterion in asymptotically (anti-)de Sitter spacetimes. Although proposed for
asymptotically flat spacetimes, such a criterion is based in the Petrov classification
scheme \cite{Petrov54} for the canonical form of the Riemann tensor. In this
classification, an Einstein manifold is assumed, i.e., 
the corresponding Ricci and metric tensors satisfy the equation $R_{\mu \nu}= \kappa g_{\mu\nu}$ with constant $\kappa$. This means that the Pirani's criterion can be used to
certify the existence of gravitation waves also in the case of asymptotically (anti--)de Sitter spacetimes.

\subsection{The canonical form of the Riemann tensor for the background spacetime}

Using the Petrov technique of Ref. \cite{Petrov54}, it is possible to put the Riemann tensor
for the background metric~\eqref{eq:background} in the following canonical matrix form
\begin{equation}
R_{\ss{AB}} =  \left(
  \begin{array}{ccc|rrr}
\alpha_1   &   \cdot      & \cdot     & \cdot & \cdot    & \cdot \\
  \cdot   &  \alpha_2    & \cdot     & \cdot   & \cdot    & \cdot\\
  \cdot    &   \cdot      &  \alpha_3 &  \cdot  & \cdot   &\cdot\\
  \hline
  \cdot   &   \cdot      & \cdot     & -\alpha_1& \cdot    & \cdot\\
  \cdot   &  \cdot     & \cdot     & \cdot   & -\alpha_2 & \cdot \\
  \cdot   &   \cdot      &  \cdot  &  \cdot  &  \cdot   & -\alpha_3  
  \end{array} \right),
\label{canonica1}
\end{equation} 
where the components of the Riemann tensor are represented in a 6-dimensional
pseudo-Euclidean space and the capital indices $A$ and $B$ assume values from 1 to 6.
The scalar quantities $\alpha_i$ are given by
\begin{equation} \label{alphas}
\alpha_1 = \frac{\Lambda_c}{3}-2\Psi_2 , \quad \alpha_2=\alpha_3=\frac{\Lambda_c}{3}+\Psi_2,
\end{equation}
where $\Lambda_c$ is the cosmological constant and $\Psi_2$ is only nonzero Weyl scalar given by Eq.~\eqref{psi2}.
Since the black brane is an asymptotically AdS spacetime, it follows that
$\sum_i \alpha_i = \Lambda_c$. 
As expected, the canonical form \eqref{canonica1} represents a Petrov type-D spacetime
and, according to the Pirani's criterion, there are no gravitational waves in such a background.

\subsection{Riemann canonical form for nonzero wave number polar
perturbations}

In this section we write the canonical Riemann tensor
for the polar perturbations of the metric~\eqref{eq:background} with an arbitrary wave number.
Again, the tetrad-gauge freedom is used to make $\delta \Psi_1=\delta \Psi_3=0$. 
In this case, the canonical form of the Riemann tensor is
  \begin{equation}
   R_{\ss{AB}} = \scriptstyle \left(
  \begin{array}{ccc|ccc}
   \alpha_1    &   \cdot  & \cdot & \cdot & \cdot    & \cdot \\
  \cdot   & \alpha_2    & \cdot  & \cdot    & \cdot   & \cdot  \\
  \cdot   &   \cdot     & \alpha_3 &  \cdot   & \cdot  &  \cdot  \\
  \hline
  \cdot    &   \cdot   & \cdot   &-\alpha_1 & \cdot  & \cdot  \\
  \cdot  &  \cdot    & \cdot   & \cdot    & -\alpha_2  & \cdot  \\
  \cdot  &   \cdot   &  \cdot  &  \cdot   &  \cdot   & -\alpha_3  
  \end{array}   \right),
\label{campo}
 \end{equation}
where the eigenvalues $\alpha_1$, $\alpha_2$, and $\alpha_3$ can be written as
\begin{equation}
\begin{split}
\alpha_1=  & \frac{\Lambda_c}{3}-2\Psi_2 -2\delta \Psi_2\,, \\
\alpha_2 =  & \frac{\Lambda_c}{3}+\Psi_2 + \delta \Psi_2 +\sqrt{\delta \Psi_0 \delta \Psi_4}\,, \\
\alpha_3=  & \frac{\Lambda_c}{3}+\Psi_2 + \delta \Psi_2 -\sqrt{\delta \Psi_0 \delta \Psi_4}\,.
\end{split}
\label{aut-valor}
\end{equation}
The Fourier transforms of the perturbations $\delta \Psi_0$ and $\delta \Psi_4$
are given respectively by Eqs.~\eqref{Psi0Pot3} and \eqref{Psi4Pot3}, 
and $\delta \Psi_2$ can be made zero by an
appropriate coordinate-gauge choice.

The canonical matrix form \eqref{campo} represents a Petrov type-I spacetime and, from the
Pirani's criterion, there is no gravitational-wave propagation in this spacetime.
This finding seems to be in opposition to the results of the analysis presented in
Sec.~\ref{zeroKpolar}. However, as discussed in Ref.~\cite{Araneda-Dotti2015},
any kind of gravitational perturbation in a Petrov type-D background leads, in general,
to a type-I spacetime. It was also shown in the
same work that metric perturbations of a Schwarzschild black hole (with multipole index 
$l\geq 2$) also lead to a Petrov type-I spacetime. Such a result contradicts the
Pirani's criterion, in view of the well-established fact that gravitational perturbations
of a Schwarzschild black hole with multipole number higher than unity describe the
propagation of gravitational waves.

The origin of the contradiction is the Pirani's assumption that the gravitational radiation
propagates along a single direction. Nevertheless, in a spacetime with nonvanishing scalars $\Psi_4$ and $\Psi_0$,
both the ingoing and outgoing waves are present. In this case, the spacetime is
more general than the ones described by the canonical forms of type II (or type N) and type III. Therefore,
the only option in the Petrov classification scheme that the problem fits in is the type I. In the next section 
the Pirani's criterion for polar perturbations with $k=0$ is discussed and, by imposing the ingoing-wave condition at the event horizon,  it is shown that the Riemann tensor reduces to
the typical form characterizing a type-II Petrov spacetime.

\subsection{Riemann canonical forms for zero wave number polar
perturbations} 
\label{sec:VB}

For the polar gravitational perturbations with zero wave numbers, the canonical 
form of the Riemann tensor and its eigenvalues are the same as presented, respectively,
in Eqs.~\eqref{campo} and~\eqref{aut-valor}. However, adopting the coordinate gauge of Ref~\cite{Miranda-Zanchin2007},
the perturbation $\delta \Psi_2$ is now given by Eq.~\eqref{Weyl-nulo2}, and it is necessary to
take $k=0$ in the expressions for the Fourier transforms of $\delta \Psi_0$ and $\delta \Psi_4$ 
given by Eqs.~\eqref{Psi0Pot3} and \eqref{Psi4Pot3}, respectively.

As discussed in Sec.~\ref{zeroKpolar}, it is interesting
to consider an ingoing-wave condition in the region close to the horizon
and to investigate the canonical form of the Riemann tensor in that limit, i.e.,
for $r_*\rightarrow -\infty$. Once again, it is more
convenient to work with the variables $Y_{+2}$ and $Y_{-2}$, instead of the
corresponding Weyl scalars $\delta \Psi_0$ and $\delta \Psi_4$,
respectively.
As shown above, $Y_{-2}$ vanishes at the horizon while $Y_{+2}$  is
given by Eq.~\eqref{Y2}. So, the canonical form of the Riemann tensor
for $r_*\rightarrow -\infty$ is given by
  \begin{equation}
   R_{\ss{AB}} = \scriptstyle \left(
  \begin{array}{ccc|ccc}
   \alpha_1    &   \cdot    & \cdot     & \cdot & \cdot   &  \cdot   \\
  \cdot   & \alpha_2 -\sigma    & \cdot  & \cdot  & \cdot   & -\sigma   \\
  \cdot  &   \cdot   & \alpha_2+\sigma  &  \cdot  & -\sigma  &  \cdot  \\
  \hline
  \cdot   &   \cdot  & \cdot   &-\alpha_1 & \cdot    & \cdot        \\
  \cdot  &  \cdot    & -\sigma    & \cdot  & -\alpha_2+\sigma  & \cdot \\
  \cdot  &   -\sigma &  \cdot  &  \cdot   &  \cdot & -\alpha_2 - \sigma  
  \end{array} \right),
\label{campo2}
 \end{equation}
where the eigenvalues are
\begin{equation}
\begin{split}
\alpha_1 =  & \frac{\Lambda_c}{3}-2\Psi_2 -2\delta \Psi_{2}\,, \\
\alpha_2 =  & \alpha_3 = \frac{\Lambda_c}{3}+\Psi_2 + \delta \Psi_{2}\,,
\end{split}
\label{aut-valor2}
\end{equation}
and
\begin{equation}
\sigma = \frac{f^{-1}}{4r}Y_{+2}
\label{tau}
\end{equation}
is the contribution of the ingoing gravitational wave.
Hence, close to the horizon, the Riemann tensor assumes the canonical form of a type-II gravitational field, which, according to the Pirani's criterion, characterizes the presence of the gravitational radiation.

Thus, the conclusion here is that gravitational perturbations whose modes describe simultaneously ingoing and outgoing gravitational radiation, i.e., perturbations with both $\delta \Psi_0$ and $\delta \Psi_4$ different from zero, generate a Riemann tensor of type I in the Petrov classification. In this case, the Pirani criterion fails to identify the gravitational waves.  On the other hand, as we have just shown, for the particular case 
when $\delta \Psi_4 = 0$,  the Riemann tensor takes the type II form and then one concludes from the Pirani criterion that it describes ingoing gravitational waves. 
This result suggests that such a criterion is conclusive as long as the radiation propagates along a specific direction in the spacetime.

\section{Zero wave number polar gravitational perturbations in the
Kodama-Ishibashi-Seto formalism}
\label{Sec:IVB}

As a final analysis, we explore here the formalism of Kodama, Ishibashi and Seto \cite{KISS2000}
to show explicitly that zero wave number polar perturbations represent gravitational waves
propagating in the radial direction. We start by setting up notation. The metric is split into the form
\begin{equation}
 ds^2 = g_{\mu \nu}dz^{\mu}dz^{\nu}=g_{ab}(y)dy^ady^b+r^2(y)d\Omega^2_{\ss{2}},
 \label{Kmetric}
\end{equation}
where $y^a = \{t,\,r \}$ and  $d\Omega^2_{\ss{2}} =
\gamma_{ij}(x)dx^idx^j$ is the metric of the two-dimensional maximally
symmetric space with coordinates $x^i = \{\varphi, \, z  \}$. 
The background metric is given by $g_{ab} = {\rm
diag}\left(-f,\,1/f\right)$ and $\gamma_{ij} = \delta_{ij}$ with 
$\delta_{ij}$ being the Kronecker delta .

For the case of polar gravitational perturbations, the metric perturbations
$h_{\mu\nu}$ may be decomposed in terms of the scalar harmonic function 
$\mathbb{S}$ as \cite{KISS2000}
\begin{equation}
\begin{split}
 & h_{ab} = f_{ab}\mathbb{S}, \\
 & h_{ai}=rf_a\mathbb{S}_i, \\
& h_{ij}=2r^2(H_L\gamma_{ij}\mathbb{S}+ H_T\mathbb{S}_{ij}),
 \label{KPertur}
\end{split}
 \end{equation}
where $\mathbb{S}$ represents the solutions of the harmonic equation $\left(\hat D_i\hat D^i + k^2\right) \mathbb{S}=0$, and quantities  $\mathbb{S}_i$ and $\mathbb{S}_{ij}$ introduced above are respectively the vectorial and tensorial harmonic functions defined by 
\begin{equation}
 \begin{split}
\mathbb{S}_i & = -\frac{1}{k}\hat{D}_i\mathbb{S}, \\  
 \mathbb{S}_{ij} & =  \frac{1}{k^2}\hat{D}_i \hat{D}_j \mathbb{S} +\frac{1}{2}\gamma_{ij}\mathbb{S},
  \end{split}
\label{harmonic}
\end{equation}
with $\hat{D}_i$ standing for the covariant derivative with respect to
the metric $\gamma_{ij}$. Again, for black-brane perturbations, the
wave number $k$ may assume any real non-negative value.

The coefficients of the decompositions \eqref{KPertur} are not invariant
under the infinitesimal gauge transformation $x^{\mu}\rightarrow x^{\mu}+\xi^{\mu}$,
i.e.,  they depend on the identification map between points of the
background spacetime and the physical spacetime.
As well as the perturbations in the metric, the vector $\xi^{\mu}$ can be decomposed
in terms of the harmonic functions as
\begin{equation}
 \xi_a = T_a \mathbb{S},  \qquad \xi_i = rL \mathbb{S}_i. 
 \label{gaugeT2}
\end{equation}

As discussed in Refs.~\cite{KISS2000,Kodama-Ishibashi2003} for $k^2>0$,
it is possible to combine the coefficients in Eqs. \eqref{KPertur}
to build the following gauge-invariant quantities:
\begin{equation}
    \begin{split}
 & F=H_L+\frac{1}{2}H_T+\frac{1}{r}D^arX_a,\\
 & F_{ab}=f_{ab}+D_aX_b+D_bX_a,
 \end{split}
\end{equation}
where $X_a=\dfrac{r}{k}f_a+D_a H_T$ and $D_a$ is the covariant derivative
with respect to the metric $g_{ab}$. However, these relations do not hold for $k^2=0$ and, moreover, as shown in Ref.~\cite{Mukohyama2000}, in such a case the functions $F$ and $F_{ab}$ are no longer gauge-invariant quantities. 
Thus, once again a detailed study of the gauge freedom of the polar perturbations with zero wave numbers is mandatory.

The first step is then to write the transformations of the quantities $H_T$, $H_L$, $f_{a}$ and $f_{ab}$, defined in \eqref{KPertur} under the infinitesimal coordinate transformation $x^{\mu}\rightarrow x^{\mu}+\xi^{\mu}$, with $\xi^{\mu}$ given by \eqref{gaugeT2}. For a vanishing wave numbers $k=0$ (besides having put $m=0$ since the beginning), it follows
\begin{equation}
\begin{aligned}[1]
  &f_{ab}  \rightarrow f_{ab}- D_aT_b-D_bT_{a}\,, \!&    H_T&  \rightarrow H_{T}\,,\\
 & f_a  \rightarrow f_a-rD_a\left(\frac{L}{r}\right), \!& H_L &
  \rightarrow H_L -\frac{D^ar}{r}T_{a}\,.  
   \end{aligned}
\label{Ktransf2}
\end{equation}
It is promptly seen that the function $H_T$ is now gauge invariant.
Additionally, convenient choices of $L$ in two successive gauge transformations
can lead $f_a$ to zero, i.e., $f_t =f_r=0$. Another gauge degree of freedom is fixed
by choosing $T_r$ so as to get rid of the perturbation $H_L$, i.e., we also may choose $H_L=0$.
At last, the function $T_t$ helps to take the $2\times 2$ matrix $f_{ab}$ into a
diagonal form, i.e., with $f_{rt}=f_{tr}=0$. These choices fix all the
gauge degrees of freedom, reducing the perturbed metric to
\begin{equation}
 ds^2 = ( g_{ab}+f_{ab}\mathbb{S})dy^ady^b
+r^2(\gamma_{ij}+2H_T\mathbb{S}_{ij})dx^idx^j,
 \label{Kmetric2}
\end{equation}
where $g_{ab}$ and $\gamma_{ij}$ stand for the background
values of the metric, cf. Eq.~\eqref{Kmetric}.

Now let us examine the harmonic functions.
According to Kodama and Ishibashi \cite{Kodama-Ishibashi2003}, for $k=0$ the scalar harmonic $\mathbb{S}$
is a constant and the vectorial and tensorial harmonic functions in \eqref{harmonic} are
not defined. On the other hand, Dias and Reall assume in Ref.~\cite{Dias20132} that,
for vanishing wave numbers, $\mathbb{S}_i$ and $\mathbb{S}_{ij}$ are zero by definition.
Therefore, in both papers, $\mathbb{S}_{ij}$ is neglected and it is argued that
the metric \eqref{Kmetric2} preserves the symmetry of the background and, on the basis
of the Birkhoff theorem, the perturbation $f_{ab}$ represents just a variation in the mass
parameter of the black brane. This is clearly true for 
spherically symmetric black-hole spacetimes, but it contradicts the
results presented in the preceding section regarding the zero wave number polar perturbations of black branes.

To solve this seeming inconsistency, we first notice that the harmonic functions
$\mathbb{S}_{i}$ and $\mathbb{S}_{ij}$ in Eq.~\eqref{harmonic} are not defined for $k=0$.
In order to raise the indeterminacy, we follow the standard procedure by calculating
explicitly such functions for arbitrary $k$ and by taking the limit of each function as $k$ goes
to zero. The scalar harmonic $\mathbb{S}$ for a black-brane spacetime is of the following form
\begin{equation}
 \mathbb{S}= e^{\pm i\,{k_j x^j}},
 \label{harmonicS}
 \end{equation}
with $x^i=(\varphi,z)$ and $k^i=(0,k)$, and where we have taken $k^\varphi =0$ because we are dealing with axisymmetric perturbations. For the planar geometry of the black-brane spacetimes, the covariant derivatives in Eqs.~\eqref{harmonic} reduce to partial derivatives, and we get
\begin{equation}
\begin{split} \label{harmonicT}
 \mathbb{S}_{zz} & =  \frac{1}{k^2}\hat{\partial}_z \hat{\partial}_z
(e^{\pm ik z})
 +\frac{1}{2}\gamma_{zz}\,e^{\pm ik z} = -\frac{1}{2} e^{\pm i k z}, \\
 \mathbb{S}_{\varphi \varphi} & = \frac{1}{2}\gamma_{\varphi
\varphi}\,e^{\pm i k z} = \frac{1}{2} e^{\pm i k z}, \\
  \mathbb{S}_{z \varphi} & =  \mathbb{S}_{\varphi z} = 0.
\end{split}
 \end{equation}
Notice that we have neglected the vectorial harmonic functions $\mathbb{S}_{i}$ because they do
not appear in the final perturbed metric \eqref{Kmetric2}.
Finally, the conclusion is that the limit $k\rightarrow 0$ of
Eqs.~\eqref{harmonicT} is well defined and gives 
\begin{equation}
 \mathbb{S}\rightarrow 1,\quad \ \
 \mathbb{S}_{\varphi \varphi} \rightarrow \frac{1}{2},\quad \ \
 \mathbb{S}_{zz}\rightarrow -\frac{1}{2}.\ \
\end{equation}

Collecting the foregoing results and substituting into the metric \eqref{Kmetric2},
it follows 
 \begin{equation}
  ds^2 = (g_{ab}+f_{ab})dy^ady^b+r^2(e^{H_T}d\varphi^2+e^{-H_T}dz^2),
  \label{metricK3}
 \end{equation}
where we have used the identity $e^{\pm H_T}=1\pm H_T$, which holds for
first-order perturbations. Furthermore, using the Einstein equations we find
that
\begin{equation}
-(g_{tt}+f_{tt})= (g_{rr}+f_{rr})^{-1}=f(r,M+\delta M)    
\end{equation}
Comparing the present approach to the formulation of Sec. \ref{zerownpolar}, we find the
relation $H_T=2\psi$. This shows that metric \eqref{metricK3} is
identical to that shown in Eq.~\eqref{Metri-null} and obtained in
Ref.~\cite{Miranda-Zanchin2007} by means of the Chandrasekhar
gauge formalism. This result confirms that zero wave number polar metric perturbations
represent also gravitational waves.

\section{Discussion}
\label{sec:V}

We have reviewed the physical interpretation of the gravitational
perturbations of AdS black branes. The focus was on the
identification of a given metric variation with the presence of gravitational waves.
The chief motivation is the conflict on the interpretation of the zero wave number
polar perturbations of black branes (cf. Refs. \cite{Kodama-Ishibashi2003}
and \cite{Miranda-Zanchin2007}). In particular, we have just shown that such
perturbations in fact represent gravitational waves propagating in the
radial direction, a situation that happens for black branes but not for
spherically symmetric black holes. 

We started by setting up the necessary basic formulation and by
showing that the equations in the Chandrasekhar gauge formalism accommodate
both the nonzero and the zero wave number gravitational fluctuations.
The explicit form of the perturbation equations for the zero wave number cases
has been needed for such an analysis. 
For the polar sector, this task was accomplished by writing
the curvature perturbations in terms of the quantities
$V^{\ss{(+)}}$, $W^{\ss{(+)}}$ and $\widetilde{Z}^{\ss{(+)}}$ [cf. Eqs.~\eqref{Pot+},
\eqref{Z+}, \eqref{W+}],  and then by taking the limit of vanishing wave numbers. 
This approach for polar perturbations of planar black holes with $k=0$ is quite different from the approach for spherically symmetric black holes, since it is not possible to investigate the special modes $l=0$ and $l=1$ from the general equations of perturbations with $l \geq 2$ in the spherically symmetric case.

In the analysis of the gravitational perturbations
we performed a direct evaluation of the complex Weyl scalars in terms of the perturbations in the metric.
The resulting expressions are in complete accord with those obtained for the vanishing wave number case in Ref. \cite{Miranda-Zanchin2006}
via the Newman-Penrose formalism and the Chandrasekhar transformation theory.
Thereby, the equivalence between both procedures, which was known to hold for spherical Schwarzschild black holes, was extended for black branes.  

As an additional technique, we relied on the work by Pirani \cite{Pirani57} to study the physical meaning of the polar-sector
fluctuations. Using this technique, we showed that perturbations with both vanishing and  nonvanishing wave numbers lead to Petrov type-I spacetimes. However, according to the Pirani's criterion, this type of spacetime
would not be associated to the propagation of gravitational waves. Obviously, this criterion presents problems in some cases. For instance, the ondulatory character of nonvanishing wave number perturbations of black branes is well established in the literature. Moreover, as shown in Ref.~\cite{Araneda-Dotti2015},  non-stationary black-hole perturbations of the  Kerr-Newman family lead
to Petrov type-I spacetimes, a result which confirms that Pirani's criterion is not conclusive when dealing with Petrov type-I spacetimes. 

In the analysis of perturbations with vanishing wave numbers by using the approach by Pirani, we performed an additional investigation. 
Our results show that such perturbations correspond to type-II gravitational fields
in the region very close to the horizon just after imposing the condition of
no outgoing waves in that neighborhood. According to the Pirani's criterion,
this kind of gravitational field characterizes the presence of gravitational waves.
Again, this outcome is in
agreement with the study of Ref.~\cite{Araneda-Dotti2015}, where it was
shown that in the particular case of vanishing $\delta \Psi_0$ or $\delta \Psi_4$,
gravitational perturbations in a type-D background lead to type-II spacetimes. From the analysis of these results we conclude that the Pirani criterion is not conclusive for general perturbations of black branes. The reason is that such a criterion neglects the possibility of coexisting ingoing and outgoing waves and, therefore, the Szerekes proposal is a more efficient invariant technique to investigate the existence of gravitational waves that propagate in different principal null directions.

In regard to the Kodama-Ishibashi-Seto \cite{KISS2000} gauge-invariant formalism,
we have verified that polar gravitational perturbations of the black branes have a
well-defined behavior in the limit of vanishing wave numbers, also in complete agreement with the
presence of gravitational waves propagating along the radial direction.
It is worth mentioning here that the main aim of the work of Ref.~\cite{Kodama-Ishibashi2003}
was to investigate the metric linear fluctuations in higher-dimensional spacetimes
and the analysis of the vanishing wave number perturbations of black branes was made {\it en passant}, in a short comment within
a long paper presenting many interesting results.

As a possible extension of the present work, it would be interesting to use the physical
interpretation of the higher-dimensional Weyl scalars~\cite{Podolsky-Svarc2012}
to extract the meaning of the gravitational perturbations of a black brane
in a general $d-$dimensional spacetime. This is a work in progress by ourselves.

\section*{Acknowledgments}

This work is partly supported by Funda\c{c}\~ao de Amparo \`a Pesquisa do Estado de S\~ao Paulo (FAPESP), Brazil, Grant No. 2015/26858-7. 
We also thank partial financial support from  Conselho Nacional de Desenvolvimento Cient\'\i fico e Tecnol\'ogico (CNPq), Brazil, Grants  No.~308346/2015-7 and No.~309609/2018-6, and from Coordena\c{c}\~ao de
Aperfei\c{c}oamento do Pessoal de N\'\i vel Superior (CAPES), Brazil, Grants 
No.~88881.064999/2014-01 and No.~88881.310352/2018-01. A. S. M. thanks financial support from Fundação de Amparo à Pesquisa do Estado da Bahia (FAPESB), Brazil.

\bibliography{referencias}

\end{document}